# Confidence Interval of Probability Estimator of Laplace Smoothing

Masato Kikuchi, Mitsuo Yoshida, Masayuki Okabe and Kyoji Umemura
Department of Computer Science and Engineering
Toyohashi University of Technology
Toyohashi, Japan

*Abstract*—Sometimes, we do not use a maximum likelihood estimator of a probability but it's a smoothed estimator in order to cope with the zero frequency problem. This is often the case when we use the Naive Bayes classifier. Laplace smoothing is a popular choice with the value of Laplace smoothing estimator being the expected value of posterior distribution of the probability where we assume that the prior is uniform distribution. In this paper, we investigate the confidence intervals of the estimator of Laplace smoothing. We show that the likelihood function for this confidence interval is the same as the likelihood of a maximum likelihood estimated value of a probability of Bernoulli trials. Although the confidence interval of the maximum likelihood estimator of the Bernoulli trial probability has been studied well, and although the approximate formulas for the confidence interval are well known, we cannot use the interval of maximum likelihood estimator since the interval contains the value 0, which is not suitable for the Naive Bayes classifier. We are also interested in the accuracy of existing approximation methods since these approximation methods are frequently used but their accuracy is not well discussed. Thus, we obtain the confidence interval by numerically integrating the likelihood function. In this paper, we report the difference between the confidence interval that we computed and the confidence interval by approximate formulas. Finally, we include a URL, where all of the intervals that we computed are available.

I. INTRODUCTION

Thi Suppose we need to classify documents using the Naive Bayes classifier. Usually, the probability that a document belongs to a $class_j$ is modeled as:

$$p(doc \mid class_j) = \prod_i p(word_i \mid class_j),$$

where $p(word_i \mid class_j)$ is the probability that a $word_i$ belongs to the class. The maximum likelihood estimated value $\hat{p}(word_i \mid class_j)$ is expressed as:

$$\hat{p}(word_i \mid class_j) = \frac{T(class_j, word_i)}{\sum_{word' \in V} T(class_j, word')},$$

where $T(class_j, word_i)$ is the number of occurrences of the $word_i$ in the class and $V$ is a set of all the words included in the training data. When a maximum likelihood estimator is used, $\hat{p}(word_i \mid class_j)$ is 0 where $\exists word_i \notin V$. Then, $\hat{p}(doc \mid class_j)$ that is the product of $\hat{p}(word_i \mid class_j)$ is also 0 and that other words are ignored. This means that if there is one word that does not appear in training set, the document cannot be classified. This is called a zero frequency problem. In this case, $p(word_i \mid class_j)$ would be small but should not be estimated as 0.

To cope with this problem, smoothing is usually used. Methods of smoothing are various. For example, the estimated value $\hat{p}(word_i \mid class_j)$ by Laplace smoothing is expressed as:

$$\hat{p}^*(word_i \mid class_j) = \frac{T(class_j, word_i) + 1}{\sum_{word' \in V} \{T(class_j, word') + 1\}},$$

where $\hat{p}^*(word_i \mid class_j)$ is the value that is assuming that all the words appear once by adding 1 to the frequency of the words.

Now, we discuss the 2-class identification problem because it is a simple case. We consider the event that the Bernoulli trials where success probability is $\theta (\in [0,1])$ succeed $x$ in $n$ times. The probability $p(x \mid \theta, n)$ that this event occurs is expressed as:

$$p(x \mid \theta, n) = {}_nC_x \theta^x (1-\theta)^{n-x}.$$

By actually conducting trials, it is assumed that $x$ times success has been observed. Likelihood $L(\theta; n, x)$ of $\theta$ is:

$$L(\theta; n, x) = {}_nC_x \theta^x (1-\theta)^{n-x}.$$

Then, a maximum likelihood estimated value is given by the following.

$$\hat{\theta} = \arg\max_\theta L(\theta; n, x) = \frac{x}{n}.$$

When estimating the appearance probability of words, this is expressed as $\hat{p}(word_i \mid class_j)$. According to the Bayesian framework, let the prior distribution of $\theta$ be a uniform distribution $\pi(\theta)$ from 0 to 1. Then, let the posterior distribution of $\theta$ be $p(\theta \mid n, x)$ when succeeding $x$ in $n$ times. $p(\theta \mid n, x)$ is obtained as:

$$p(\theta \mid n, x) = \frac{p(x \mid \theta, n)\pi(\theta)}{\int p(x \mid \theta', n)\pi(\theta')d\theta'}.$$

When the distribution is $p(\theta|n,x)$, the expected value $\bar{\theta}$ of $\theta$ is expressed as follows.

$$\bar{\theta} = \int \theta \cdot p(\theta|n,x) d\theta = \frac{x+1}{n+2}.$$

This is known to be the probability estimator of Laplace smoothing. When estimating the appearance probability of words, this is expressed as $\hat{p}^*(word_i | class_j)$.

Now, we consider the problem that estimating the confidence interval of $\bar{\theta}$. The denominator does not depend on $\theta$, and $\pi(\theta)$ also does not depend on $\theta$ within $0 \leq \theta \leq 1$. Therefore, the following formula is established within $0 \leq \theta \leq 1$.

$$p(\theta|n,x) = \frac{p(x|\theta,n)\pi(\theta)}{\int p(x|\theta',n)\pi(\theta')d\theta'} \propto p(x|\theta,n) = L(\theta;n,x).$$

This means that the likelihood for the confidence interval of $\bar{\theta}$ is the same as the likelihood for the maximum likelihood estimated value by the Bernoulli trials except for the constant factor. The problem in estimating the confidence interval of the maximum likelihood estimated value has been researched in a number of articles, and approximate formulas for the confidence interval are used. There are many kinds of formulas such as the formula using normal approximation, Wilson's formula [1], Agresti & Coull's formula [2], Clopper & Pearson's formula [3], Sterne's formula [4], Crow's formula [5], Blyth & Still's formula [6] and Blaker's formula [7]. Then, numerical comparisons of the confidence intervals by these formulas, proposals of new formulas and so on have been undertaken [8] [9]. However, these formulas may not be able to approximate, under conditions that have an important meaning by smoothing, that is, in the case where the probability $p$ to estimate is close to 0 or $n$ is small. Moreover, because the confidence interval may contain the 0 value, the zero frequency problem occurs when using the confidence interval. It is natural that the confidence interval of the maximum likelihood estimator $\hat{\theta}$ includes 0 because the estimator can be 0. For the confidence interval of expectation $\bar{\theta}$, the confidence interval should not include 0.

The difference of the integral method between the confidence interval of the probability estimated value and the confidence interval of the probability estimated value for smoothing is shown in Fig. 1. As far as we know, the confidence interval of the expectation has not been reported. Therefore, we have computed the confidence interval of the probability estimated value for smoothing by integrating the above likelihood function $L(\theta;n,x)$. There are previous works that utilize the confidence interval for the Naïve Bayes classifier [10] [11]. However, these works do not use $\bar{\theta}$ but $\hat{\theta}$, and do not pay attention to zero frequency.

In this paper, we obtain the confidence interval by numerically integrating the likelihood function and report the difference in values between the confidence interval we obtained and the confidence interval calculated by approximate formulas. Finally, we include a URL, where all of the intervals that we computed are available, as its size is too large to be included in this paper.

## II. ESTIMATING METHOD OF THE CONFIDENCE INTERVAL

We describe a method using approximate formulas and a method using numerical integral as estimating method for the confidence interval of the probability estimated value.

### II-A. Approximate formulas for the confidence interval

The formula using normal approximation is the most well used formula as a formula of the confidence interval. The $100(1-\alpha)\%$ confidence interval using normal approximation is expressed as:

$$\hat{p} - z_{\alpha/2}\sqrt{\frac{\hat{p}(1-\hat{p})}{n}} \leq p \leq \hat{p} + z_{\alpha/2}\sqrt{\frac{\hat{p}(1-\hat{p})}{n}},$$

where $\hat{p} := x/n$, and $z_{\alpha/2}$ denotes the $100(1-\alpha/2)\%$ point of the standard normal distribution. When using this formula, the following conditions should be satisfied for approximating the binomial distribution by the normal distribution [12].

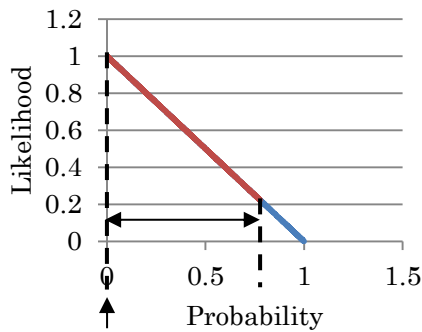

$\hat{\theta} = 0$ : maximum likelihood estimator

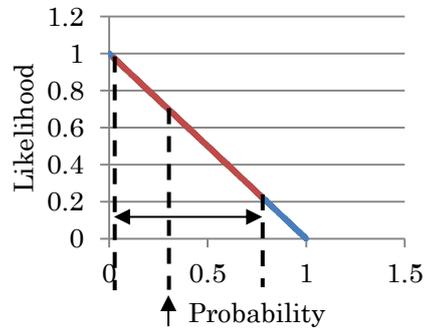

$\bar{\theta} = 1/3$ : expected value of the probability

Figure 1. Although the confidence interval of $\hat{\theta}$ should include the maximum likelihood estimated value 0, the value of 0 will cause the zero frequency problem. On the other hand, the confidence interval of $\bar{\theta}$ should not to include 0 because 0 is an extreme value.

(1) $(1-p)$ are $\geq 5$ (or 10);
(2) $np(1-p)$ are $\geq 5$ (or 10);
(3) $n\hat{p}, n(1-\hat{p})$ are $\geq 5$ (or 10);
(4) $\hat{p} \pm 3\sqrt{\hat{p}(1-\hat{p})/n}$ does not contain 0 or 1;
(5) $n$ quite large;
(6) $n \geq 50$ unless $p$ is very small.

Therefore, when the above conditions are not satisfied, Clopper & Pearson's "exact" formula is known as the formula to obtain the confidence interval [13] [14]. Clopper & Pearson's $100(1-\alpha)\%$ confidence interval is expressed as:

$$\frac{v_1}{v_1 + v_2 F_{\alpha/2}(v_1, v_2)} \leq p \leq \frac{v_3 F_{\alpha/2}(v_3, v_4)}{v_4 + v_3 F_{\alpha/2}(v_3, v_4)},$$

where $v_1 := 2(n-x+1)$, $v_2 := 2x$, $v_3 := 2(x+1)$ and $v_4 := 2(n-x)$. $F_{\alpha/2}(v_1, v_2)$ denotes the $100(1-\alpha/2)\%$ point of the F-distribution with degrees of freedom $(v_1, v_2)$ and $F_{\alpha/2}(v_3, v_4)$ denotes the $100(1-\alpha/2)\%$ point of the F-distribution with degrees of freedom $(v_3, v_4)$. When $n$ is small, this formula takes a larger confidence interval than the true confidence interval [2].

*II-B. The confidence interval by numerical integral*

The confidence interval of the probability estimated value $\bar{p}$ for smoothing can be obtained by numerically integrating the likelihood function $L(p;n,x)$ of the binomial distribution. $L(p;n,x)$ is expressed as:

$$L(p;n,x) = {}_nC_x p^x (1-p)^{n-x}.$$

When $L(p;n,x)$ satisfies the following relationship:

$$\int_0^{p_{lb}} L(p;n,x)dp = \int_{1-p_{ub}}^1 L(p;n,x)dp = \frac{\alpha}{2}\int_0^1 L(p;n,x)dp,$$

the $100(1-\alpha)\%$ confidence interval by numerical integral is expressed as:

$$p_{lb} \leq p \leq p_{ub}.$$

In this paper, we use Simpson's rule to integrate $L(p;n,x)$ numerically. For example, let a function be $f(y)$. Calculate the integral of $f(y)$ in the range $[a,b]$. First, $[a,b]$ is divided into $k$ sections to be the width $h = (b-a)/k$. Next, the area is approximated by the secondary curve passing each of the three division points of $f(y)$. Finally, the entire area is computed by the sum of these areas. Now, Simpson's rule is expressed as:

$$\int_a^b f(y)dy \approx \frac{h}{3}\left[f(y_0) + 2\sum_{i=1}^{k/2-1} f(y_{2i}) + 4\sum_{i=1}^{k/2} f(y_{2i-1}) + f(y_k)\right].$$

For this reason, numerical integration of $L(p;n,x)$ is computed with high accuracy by using the GNU Multiple Precision Arithmetic Library (GMP). The larger $k$, the number of sections to divide $[a,b]$ becomes, the more Simpson's rule improves the accuracy of the numerical integral. We discuss the accuracy of the numerical integral in Section III.

III. EXPERIMENTS

In the experiment, we compare the confidence interval by approximate formulas and the confidence interval by numerical integral numerically in Section III-A. Here, we compare the difference of values between the confidence interval by numerical integral as the theoretical value and the confidence interval by approximate formulas. We examine the accuracy of the confidence interval by numerical integral and indicate that the comparison results shown in Section III-A are correct in Section III-B.

*III-A. Comparison of the confidence intervals obtained by each methods numerically*

We compare the confidence interval by approximate formulas and the confidence interval by numerical integral numerically. We use the formula using normal approximation and Clopper & Pearson's formula as approximate formulas. $k$, the number of sections to equally divide integration range $[a,b]$ is $1048576 (= 2^{20})$. The number of trials $n$ is 5 and 1000. The number of successes $x$ is 0 to 5 when $n$ is 5 and $x$ is 0, 500, and 1000 when $n$ is 1000. Here, we choose small $n$ and large $n$ so that the difference between approximate formulas and numerical integral should become clear. The selected confidence coefficients are 95% and 99%. First, we compare the lower limit values and upper limit values in accuracy of 5 decimal places for the confidence interval by approximate formulas and the confidence interval by numerical integral. Next, the confidence interval by numerical integral is regarded as the theoretical value, and we calculate the error percentage of the confidence interval by approximate formulas.

The lower and upper limit values of the 95% confidence interval we obtained by each method are shown in Table I. When $x$ is 0, the lower limit value of the confidence interval using normal approximation is 0 and the lower limit value of Clopper & Pearson's confidence interval is also 0. When $x$ is equal to $n$, the upper limit value of the confidence interval using normal approximation is 1.0 and the upper limit value of Clopper & Pearson's confidence interval is also 1.0. These confidence intervals contain the values that are not suitable for smoothed estimator. On the other hand, our confidence interval by numerical integral does not contain 0 or 1 when $x$ is 0 or $n$. The calculation results of the error percentage of 95% confidence interval by approximate formulas with the exception of Indicated values with an underscore in Table I are shown in Table II and Table III. The error percentage of confidence interval using normal approximation is more than 5% when $n$ is 5, and more than 0.01% when $n$ is 1000. The error percentage of Clopper & Pearson's confidence interval becomes smaller as $x$ increases. It also should be noted that Clopper & Pearson's confidence interval is always wider than our interval.

The lower and upper limit values of the 99% confidence interval we obtained by each method are shown in Table IV. The calculation results of the error percentage of 99% confidence interval by approximate formulas are shown in Table V and Table VI. These results have numerical trend similar to the results for the 95% confidence interval.

TABLE I. LOWER AND UPPER LIMIT VALUES OF THE 95% CONFIDENCE INTERVAL

| n | x | Numerical integral | | Normal approximation | | Clopper & Pearson | |
|---|---|---|---|---|---|---|---|
| | | Lower limit | Upper limit | Lower limit | Upper limit | Lower limit | Upper limit |
| 5 | 0 | 0.00421 | 0.45925 | 0.00000 | 0.00000 | 0.00000 | 0.52181 |
| 5 | 1 | 0.04327 | 0.64123 | -0.15061 | 0.55061 | 0.00505 | 0.71641 |
| 5 | 2 | 0.11811 | 0.77722 | -0.02941 | 0.82941 | 0.05274 | 0.85336 |
| 5 | 3 | 0.22277 | 0.88188 | 0.17058 | 1.02941 | 0.14663 | 0.94725 |
| 5 | 4 | 0.35876 | 0.95672 | 0.44938 | 1.15061 | 0.28358 | 0.99494 |
| 5 | 5 | 0.54074 | 0.99578 | 1.00000 | 1.00000 | 0.47818 | 1.00000 |
| 1000 | 0 | 0.00002 | 0.00367 | 0.00000 | 0.00000 | 0.00000 | 0.00368 |
| 1000 | 500 | 0.46906 | 0.53093 | 0.46900 | 0.53099 | 0.46854 | 0.53145 |
| 1000 | 1000 | 0.99632 | 0.99997 | 1.00000 | 1.00000 | 0.99631 | 1.00000 |

TABLE II. ERROR PERCENTAGE OF THE 95% CONFIDENCE INTERVAL (NORMAL APPROXIMATION)

| n | x | Lower limit | | | n | x | Upper limit | | |
|---|---|---|---|---|---|---|---|---|---|
| | | Numerical integral | Normal approximation | Error [%] | | | Numerical integral | Normal approximation | Error [%] |
| 5 | 3 | 0.22277 | 0.17058 | -23.428 | 5 | 1 | 0.64123 | 0.55061 | -14.131 |
| 5 | 4 | 0.35876 | 0.44938 | 25.258 | 5 | 2 | 0.77722 | 0.82941 | 6.715 |
| 1000 | 500 | 0.46906 | 0.46900 | -0.011 | 1000 | 500 | 0.53093 | 0.53099 | 0.010 |

TABLE III. ERROR PERCENTAGE OF THE 95% CONFIDENCE INTERVAL (CLOPPER & PEARSON)

| n | x | Lower limit | | | n | x | Upper limit | | |
|---|---|---|---|---|---|---|---|---|---|
| | | Numerical integral | Clopper & Pearson | Error [%] | | | Numerical integral | Clopper & Pearson | Error [%] |
| 5 | 1 | 0.04327 | 0.00505 | -88.327 | 5 | 0 | 0.45925 | 0.52181 | 13.622 |
| 5 | 2 | 0.11811 | 0.05274 | -55.344 | 5 | 1 | 0.64123 | 0.71641 | 11.724 |
| 5 | 3 | 0.22277 | 0.14663 | -34.179 | 5 | 2 | 0.77722 | 0.85336 | 9.797 |
| 5 | 4 | 0.35876 | 0.28358 | -20.955 | 5 | 3 | 0.88188 | 0.94725 | 7.412 |
| 5 | 5 | 0.54074 | 0.47818 | -11.569 | 5 | 4 | 0.95672 | 0.99494 | 3.994 |
| 1000 | 500 | 0.46906 | 0.46854 | -0.109 | 1000 | 0 | 0.00367 | 0.00368 | 0.100 |
| 1000 | 1000 | 0.99632 | 0.99631 | -0.000 | 1000 | 500 | 0.53093 | 0.53145 | 0.096 |

TABLE IV. LOWER AND UPPER LIMIT VALUES OF THE 99% CONFIDENCE INTERVAL

| n | x | Numerical integral | | Normal approximation | | Clopper & Pearson | |
|---|---|---|---|---|---|---|---|
| | | Lower limit | Upper limit | Lower limit | Upper limit | Lower limit | Upper limit |
| 5 | 0 | 0.00083 | 0.58648 | 0.00000 | 0.00000 | 0.00000 | 0.65342 |
| 5 | 1 | 0.01872 | 0.74600 | -0.26152 | 0.66152 | 0.00100 | 0.81490 |
| 5 | 2 | 0.06627 | 0.85640 | -0.16524 | 0.96524 | 0.02288 | 0.91717 |
| 5 | 3 | 0.14359 | 0.93372 | 0.03475 | 1.16524 | 0.08282 | 0.97711 |
| 5 | 4 | 0.25399 | 0.98127 | 0.33847 | 1.26152 | 0.18509 | 0.99899 |
| 5 | 5 | 0.41351 | 0.99916 | 1.00000 | 1.00000 | 0.34657 | 1.00000 |
| 1000 | 0 | 0.00000 | 0.00527 | 0.00000 | 0.00000 | 0.00000 | 0.00528 |
| 1000 | 500 | 0.45937 | 0.54062 | 0.45920 | 0.54079 | 0.45885 | 0.54114 |
| 1000 | 1000 | 0.99472 | 0.99999 | 1.00000 | 1.00000 | 0.99471 | 1.00000 |

TABLE V. ERROR PERCENTAGE OF THE 99% CONFIDENCE INTERVAL (NORMAL APPROXIMATION)

| n | x | Lower limit | | | n | x | Upper limit | | |
|---|---|---|---|---|---|---|---|---|---|
| | | Numerical integral | Normal approximation | Error [%] | | | Numerical integral | Normal approximation | Error [%] |
| 5 | 3 | 0.14359 | 0.03475 | -75.799 | 5 | 1 | 0.74600 | 0.66152 | -11.324 |
| 5 | 4 | 0.25399 | 0.33847 | 33.261 | 5 | 2 | 0.85640 | 0.96524 | 12.709 |
| 1000 | 500 | 0.45937 | 0.45920 | -0.035 | 1000 | 500 | 0.54062 | 0.54079 | 0.030 |

TABLE VI. ERROR PERCENTAGE OF THE 99% CONFIDENCE INTERVAL (CLOPPER & PEARSON)

| n | x | Lower limit | | | n | x | Upper limit | | |
|---|---|---|---|---|---|---|---|---|---|
| | | Numerical integral | Clopper & Pearson | Error [%] | | | Numerical integral | Clopper & Pearson | Error [%] |
| 5 | 1 | 0.01872 | 0.00100 | -94.647 | 5 | 0 | 0.58648 | 0.65342 | 11.414 |
| 5 | 2 | 0.06627 | 0.02288 | -65.477 | 5 | 1 | 0.74600 | 0.81490 | 9.235 |
| 5 | 3 | 0.14359 | 0.08282 | -42.317 | 5 | 2 | 0.85640 | 0.91717 | 7.095 |
| 5 | 4 | 0.25399 | 0.18509 | -27.124 | 5 | 3 | 0.93372 | 0.97711 | 4.647 |
| 5 | 5 | 0.41351 | 0.34657 | -16.188 | 5 | 4 | 0.98127 | 0.99899 | 1.805 |
| 1000 | 500 | 0.45937 | 0.45885 | -0.113 | 1000 | 0 | 0.00527 | 0.00528 | 0.107 |
| 1000 | 1000 | 0.99472 | 0.99471 | -0.000 | 1000 | 500 | 0.54062 | 0.54114 | 0.096 |

## III-B. The accuracy of the confidence interval by numerical integral

In Section III-A, we discussed the error of 0.01%. To make this discussion meaningful, we should confirm the accuracy of confidence intervals of numerical integral. To verify the accuracy of the numbers, we double $k$, the number of sections in Simpson's rule and examine the number of digits whose values for the confidence interval are unchanged. Then, we judge that the values corresponding to the number of digits unchanged are correct. Furthermore, because we use GMP, we consider that the accuracy is determined by the number of divisions. $k$ to examine the accuracy is $1048576 \ (=2^{20})$ same as the experiment in the previous section. We examine the values of the confidence interval to eight decimal places.

The results of examining the accuracy of the 95% confidence interval and 99% confidence interval by numerical integral are shown Table VII and Table VIII respectively. Underlined parts of each table denote the digits where the values change when $k$ is doubled. The digits where the values change are sixth digits after the decimal point. For this reason, the confidence intervals of numerical integral are accurate in five or more decimal places. Moreover, the confidence intervals of numerical integral are sufficiently accurate to discuss differences even when $n$ is large enough ($n=1000$) in Section III-A.

TABLE VII.
ACCURACY OF THE 95% CONFIDENCE INTERVAL BY NUMERICAL INTEGRAL

| $n$ | $x$ | Lower limit | | Upper limit | |
| --- | --- | --- | --- | --- | --- |
| | | $k:2^{20}$ | $k:2^{21}$ | $k:2^{20}$ | $k:2^{21}$ |
| 5 | 0 | 0.00421047 | 0.00421094 | 0.45925807 | 0.45925807 |
| 5 | 1 | 0.04327201 | 0.04327201 | 0.64123439 | 0.64123487 |
| 5 | 2 | 0.11811733 | 0.11811733 | 0.77722167 | 0.77722215 |
| 5 | 3 | 0.22277832 | 0.22277784 | 0.88188266 | 0.88188266 |
| 5 | 4 | 0.35876560 | 0.35876512 | 0.95672798 | 0.95672798 |
| 5 | 5 | 0.54074192 | 0.54074192 | 0.99578952 | 0.99578905 |
| 1000 | 0 | 0.00002574 | 0.00002527 | 0.00367832 | 0.00367832 |
| 1000 | 500 | 0.46906375 | 0.46906328 | 0.53093624 | 0.53093671 |
| 1000 | 1000 | 0.99632167 | 0.99632167 | 0.99997425 | 0.99997472 |

TABLE VIII.
ACCURACY OF THE 99% CONFIDENCE INTERVAL BY NUMERICAL INTEGRAL

| $n$ | $x$ | Lower limit | | Upper limit | |
| --- | --- | --- | --- | --- | --- |
| | | $k:2^{20}$ | $k:2^{21}$ | $k:2^{20}$ | $k:2^{21}$ |
| 5 | 0 | 0.00083446 | 0.00083494 | 0.58648204 | 0.58648157 |
| 5 | 1 | 0.01872062 | 0.01872062 | 0.74600696 | 0.74600744 |
| 5 | 2 | 0.06627941 | 0.06627893 | 0.85640430 | 0.85640430 |
| 5 | 3 | 0.14359569 | 0.14359569 | 0.93372058 | 0.93372106 |
| 5 | 4 | 0.25399303 | 0.25399255 | 0.98127937 | 0.98127937 |
| 5 | 5 | 0.41351795 | 0.41351842 | 0.99916553 | 0.99916505 |
| 1000 | 0 | 0.00000476 | 0.00000524 | 0.00527858 | 0.00527906 |
| 1000 | 500 | 0.45937061 | 0.45937013 | 0.54062938 | 0.54062986 |
| 1000 | 1000 | 0.99472141 | 0.99472093 | 0.99999523 | 0.99999475 |

## IV. CONCLUSION

Since the confidence interval of the probability estimated value for smoothing $\bar{\theta}$ should not include 0, we obtain the confidence interval of $\bar{\theta}$ by integrating the likelihood function of the binomial distribution numerically. Moreover, we obtain the differences among the confidence intervals of popular approximation formulas and our computation.

In the experiment in Section III-A, we compare the confidence interval by approximate formulas and the confidence interval by numerical integral numerically. It is stated that the confidence interval by approximate formulas may include 0 or 1 value. It is also stated that the confidence interval by numerical integral does not include a value of 0 or 1. For a large $n$ of 1000, both intervals show similar value but still have slight difference. Assuming that the confidence interval by numerical integral is a theoretical value, we calculate the percentage error of the confidence interval by approximate formulas. We indicate that the confidence interval by approximate formulas is different from the confidence interval by numerical integral, having a difference of more than 0.01% even when the approximation is considered established under the condition that $n$ is large ($n=1000$). Finally, from the following URL, all of the intervals that we computed are available.

http://www.ss.cs.tut.ac.jp/CI-Laplace

APPENDIX

When $n=1$ and $x=0$, the value of each formula is as follows:

$$L(\theta;n,x) = {}_1C_0 \theta^0 (1-\theta)^{1-0} = 1-\theta,$$

$$\hat{\theta} = \arg\max_{\theta \in [0,1]} (1-\theta) = 0.$$

Its 95% confidence interval is from 0 to 0.77639 (see Fig. 1, left chart).

$$p(x \mid \theta, n) = {}_1C_0 \theta^0 (1-\theta)^{1-0} = 1-\theta,$$

$$p(\theta \mid n,x) = \frac{(1-\theta)\pi(\theta)}{\int (1-\theta')\pi(\theta')d\theta'} = \frac{(1-\theta)\pi(\theta)}{\int_0^1 (1-\theta')d\theta'} = 2(1-\theta)\pi(\theta),$$

$$\bar{\theta} = \int \theta \cdot 2(1-\theta)\pi(\theta)d\theta = \int_0^1 \theta \cdot 2(1-\theta)d\theta = \frac{1}{3}.$$

Its 95% confidence interval is from 0.0125 to 0.84188 (see Fig. 1, right chart).